# Principles of Distributed Data Management in 2020?[1]

Patrick Valduriez

INRIA and LIRMM, Montpellier – France
`Patrick.Valduriez@inria.fr`

**Abstract.** With the advents of high-speed networks, fast commodity hardware, and the web, distributed data sources have become ubiquitous. The third edition of the Özsu-Valduriez textbook *Principles of Distributed Database Systems* [10] reflects the evolution of distributed data management and distributed database systems. In this new edition, the fundamental principles of distributed data management could be still presented based on the three dimensions of earlier editions: distribution, heterogeneity and autonomy of the data sources. In retrospect, the focus on fundamental principles and generic techniques has been useful not only to understand and teach the material, but also to enable an infinite number of variations. The primary application of these generic techniques has been obviously for distributed and parallel DBMS versions. Today, to support the requirements of important data-intensive applications (e.g. social networks, web data analytics, scientific applications, etc.), new distributed data management techniques and systems (e.g. MapReduce, Hadoop, SciDB, Peanut, Pig latin, etc.) are emerging and receiving much attention from the research community. Although they do well in terms of consistency/flexibility/performance trade-offs for specific applications, they seem to be ad-hoc and might hurt data interoperability. The key questions I discuss are: What are the fundamental principles behind the emerging solutions? Is there any generic architectural model, to explain those principles? Do we need new foundations to look at data distribution?

## 1  Introduction

The 1980's were very active periods for the development of distributed relational database technology and all commercial DBMSs today are distributed. The decade of 1990's saw the development and maturation of client-server technology and the introduction of object-orientation – both as stand-alone systems and as object-relational DBMSs.

The Özsu-Valduriez textbook Principles of Distributed Database Systems [10] was first published in 1991, covering the fundamental distribution principles and techniques. The second edition was published in 1999 and included coverage of client-server systems and distributed object systems. The third edition of the book was out in April 2011. During the writing of the third edition, we have been evaluating the past and contemplating the future. It has been almost twenty years since the first edition appeared, and ten years since the second edition. As one can imagine, in a fast changing area such as this, there have been significant changes in the intervening period. As we wrote the third edition, we incorporated technologies that were developed in late 1990's and in 2000's – P2P systems, data integration, database clusters, web and XML data management, stream data management, and cloud data management. It is apparent that the last ten years have seen an accelerated investigation of distributed data management technologies spurred by advent of high-speed networks, fast commodity hardware, very heavy parallelization of hardware, and, of course, the increasing pervasiveness of the web.

---

[1] Work partially funded by the DataRing project of the French ANR.

Now, the question is what is likely to happen in the next decade; or to put it differently, if there were to be a fourth edition of our book in 2020, *what would it be? what would be new?* This is the motivation for this paper[2].

In observing the changes that have taken place over the past twenty years of our involvement with this field, what has struck as interesting is that the fundamental principles of distributed data management still hold, and distributed data management can be characterized on three dimensions: distribution, heterogeneity and autonomy of the data sources. What has changed much since and made the problems much harder, is the scale of the dimensions: very large scale distribution (cluster, P2P, web and cloud); very high heterogeneity (web); and high autonomy (web, P2P). Also interesting to note is that the fundamental principles of database fragmentation (or partitioning), data integration, transaction management, replication and relational query processing have stood the test of time. In particular, new techniques and algorithms could be presented as extensions of earlier material, using relational concepts.

Today, to support the requirements of important data-intensive applications (e.g. social networks, web data analytics), new distributed data management techniques (e.g. MapReduce, Hadoop, Peanut, Pig latin, SciDB) are emerging and receiving much attention from the research community. Although they do well in terms of consistency-flexibility-performance trade-offs for specific applications, they seem to be ad-hoc and might hurt data interoperability. The key questions are: What are the fundamental principles behind the emerging solutions? Is there any generic architectural model, to explain those principles? Do we need new foundations to look at data distribution?

In this paper, I discuss these questions. I first recall the fundamental principles of distributed data management (Section 2). Then, in Section 3, I illustrate the challenges introduced by new data-intensive applications in the context of scientific applications and cloud computing. In Section 4, I discuss emerging solutions and show that they can still be explained along the three main dimensions of distributed data management (distribution, autonomy, heterogeneity). Finally, in Section 5, I discuss what is likely to come next.

---

[2] This was also the basis for a panel at ICDE 2011 [11].

## 2   Principles of Distributed Data Management

The fundamental principle behind data management is *data independence*, which enables applications and users to share data at a high conceptual level while ignoring implementation details. This principle has been achieved by *database systems* which provide advanced capabilities such as schema management, high-level query languages, access control, automatic query processing and optimization, transactions, data structures for supporting complex objects, etc.

A *distributed database* is a collection of multiple, logically interrelated databases distributed over a computer network. A *distributed database system* is defined as the software system that permits the management of the distributed database and makes the distribution *transparent* to the users. Distribution transparency extends the principle of data independence so that distribution is not visible to users.

These definitions assume that each site logically consists of a single, independent computer. Therefore, each site has the capability to execute applications on its own. The sites are interconnected by a computer network with loose connection between sites which operate independently. Applications can then issue queries and transactions to the distributed database system which transforms them into local queries and local transactions (see Figure 1) and integrates the results. The distributed database system can run at any site s, not necessarily distinct from the data (i.e. it can be site 1 or 2 in Figure 1).

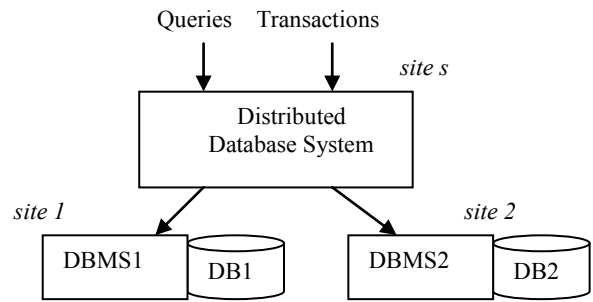

**Figure 1.** A distributed database system with two data sites

The database is physically distributed across the data sites by fragmenting and replicating the data. Given a relational database schema, for instance, fragmentation subdivides each relation into partitions based on some function applied to some tuples' attributes. Based on the user access patterns, each of the fragments may also be replicated to improve locality of reference (and thus performance) and availability. The use of a set-oriented data model (like relational) has been crucial to define fragmentation, based on data subsets.

The functions provided by a distributed database system could be those of a database system (schema management, access control, query processing, transaction support, etc). But since they must deal with distribution, they are more complex to implement. Therefore, many systems support only a subset of these functions.

When the data and the databases already exist, one is faced with the problem of providing integrated access to heterogeneous data. This process is known as *data integration*: it consists in defining a *global schema* over the existing data and *mappings* between the global schema and the local database schemas. Data integration systems have received several names such as federated database systems, multidatabase systems and, more recently, mediator systems. Standard protocols such as Open Database Connectivity (ODBC) and Java Database Connectivity (JDBC) ease data integration using SQL. In the context of the Web, mediator systems [13] allow general access to autonomous data sources (such as files, databases, documents, etc.) in read only mode. Thus, they typically do not support all database functions such as transactions and replication.

When the architectural assumption of each site being a (logically) single, independent computer is relaxed, one gets a *parallel database system* [14], i.e. a database system implemented on a tightly-coupled multiprocessor or a cluster. The main difference with a distributed database system is that there is a single operating system which eases implementation and the network is typically faster and more reliable. The objective of parallel database systems is high-performance and high-availability. High-performance (i.e. improving transaction throughput or query response time) is obtained by exploiting data partitioning and query parallelism while high-availability is obtained by exploiting replication. Again this has been made possible by the use of a set-oriented data model, which eases parallelism, in particular, independent parallelism between data subsets.

The distributed database approach has proved effective for applications that can benefit from static administration, centralized control and full-fledge database capabilities, e.g. information systems. For administrative reasons, the distributed database system typically runs on a separate server, which reduces scalability to tens of databases. Data integration systems achieve better scalability to hundreds or thousands of data sources by restricting functionality (i.e. read-only querying). Parallel database systems can also scale up to large configurations with thousands of processing nodes. However, both data integration systems and parallel database systems typically rely on a global schema that can be either centralized or replicated.

We now consider the possible ways in which a distributed DBMS may be architected. We use a classification (Figure 1) that organizes the systems as characterized with respect to three dimensions: (1) the autonomy of local systems, (2) their distribution, and (3) their heterogeneity. Autonomy, in this context, refers to the distribution of control, not of data. It indicates the degree to which individual DBMSs can operate independently. Whereas autonomy refers to the distribution (or decentralization) of control, the distribution dimension of the taxonomy deals with the physical distribution of data over multiple sites (or nodes in a parallel system). There are a number of ways DBMSs have been distributed. We distinguish between client/server (C/S) distribution and peer-to-peer (P2P) distribution (or full distribution). With C/S DBMS, sites may be clients or servers, thus with different functionality, whereas with homogeneous P2P DBMS (DDBMS in Figure 1), all sites provide the same functionality. Note that DDBMS came before C/S DBMS in the late 1970. P2P data management struck back in the 2000 with "modern" variations to deal with very large scale, autonomy and decentralized control. Heterogeneity refers to data models, query languages, and transaction management protocols. Multidatabase systems (MDBMS) deal with heterogeneity, in addition to autonomy and distribution.

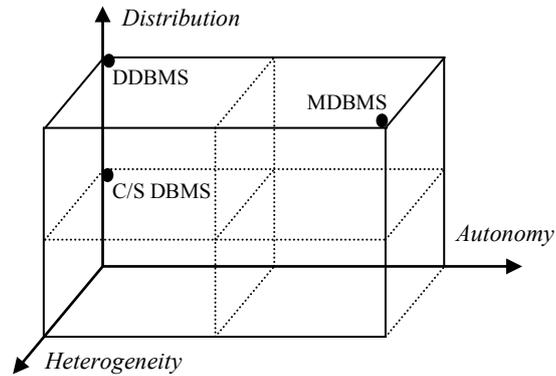

**Figure 2.** Distributed DBMS Architectures (modified after [10])

## 3  New Challenges for Distributed Data Management

The pervasiveness of the web has spurred all kinds of data-intensive applications and introduced great challenges for distributed data management. New data-intensive applications such as social networks, web data analytics and scientific applications have requirements that are not met by the traditional distributed database systems in Figure 2. What has changed much and made the problems much harder, is the scale of the dimensions: very large scale distribution, very high heterogeneity, and high autonomy. Let us illustrate the challenges for distributed data management with two important domains: scientific data management and cloud data management.

### 3.1 Scientific Data Management

Scientific data management has become a major challenge for the database and data management research community [7]. Modern science such as agronomy, bio-informatics, physics and environmental science must deal with overwhelming amounts of experimental data produced through empirical observation and simulation. Such data must be processed (cleaned, transformed, analyzed) in all kinds of ways in order to draw new conclusions, prove scientific theories and produce knowledge. However, constant progress in scientific observational instruments (e.g. satellites, sensors, large hadron collider) and simulation tools (that foster in silico experimentation, as opposed to traditional in situ or in vivo experimentation) creates a huge data overload. For example, climate modeling data are growing so fast that they will lead to collections of hundreds of exabytes expected by 2020.

Scientific data is also very complex, in particular because of heterogeneous methods used for producing data, the uncertainty of captured data, the inherently multi-scale nature (spatial scale, temporal scale) of many sciences and the growing use of imaging (e.g. satellite images), resulting in data with hundreds of attributes, dimensions or descriptors. Processing and analyzing such massive sets of complex scientific data is therefore a major challenge since solutions must combine new data management techniques with large-scale parallelism in cluster, grid or cloud environments [12].

Furthermore, modern science research is a highly collaborative process, involving scientists from different disciplines (e.g. biologists, soil scientists, and geologists working on an environmental project), in some cases from different organizations distributed in different countries. Since each discipline or organization tends to produce and manage its own data, in specific formats, with its own processes, integrating distributed data and processes gets difficult as the amounts of heterogeneous data grow.

Despite their variety, we can identify common features of scientific data [1]: massive scale; manipulated through complex, distributed workflows; typically complex, e.g. multidimensional or graph-based; with uncertainty in the data values, e.g., to reflect data capture or observation; important metadata about experiments and their provenance; heavy floating-point computation; and mostly append-only (with rare updates).

### 3.2. Cloud Data Management

Cloud computing is the latest trend in distributed computing and has been the subject of much hype. The vision encompasses on demand, reliable services provided over the Internet (typically represented as a cloud) with easy access to virtually infinite computing, storage and networking resources. Through very simple Web interfaces and at small incremental cost, users can outsource complex tasks, such as data storage, system administration, or application deployment, to very large data centers operated by cloud providers. Thus, the complexity of managing the software/hardware infrastructure gets shifted from the users' organization to the cloud provider. From a technical point of view, the grand challenge is to support in a cost-effective way the very large scale of the infrastructure which has to manage lots of users and resources with high quality of service.

However, not all data-intensive applications are good candidates for being "cloudified" [1]. To simplify, we can classify between the two main classes of data-intensive applications: On Line Transaction Processing (OLTP) and On Line Analytical Processing (OLAP). Let us recall their main characteristics. OLTP deals with operational databases of average sizes (up to a few Terabytes), is write-intensive, and requires complete ACID transactional properties, strong data protection and response time guarantees. On the other hand, OLAP deals with historical databases of very large sizes (up to Petabytes), is read-intensive and thus can accept relaxed ACID properties. Furthermore, since OLAP data are typically extracted from operational

OLTP databases, sensitive data can be simply hidden for analysis (e.g. using anonymization) so that data protection is not as crucial as in OLTP.

OLAP is more suitable than OLTP for cloud primarily because of two cloud characteristics (see the detailed discussion in [1]): elasticity and security. To support elasticity in a cost-effective way, the best solution that most cloud providers adopt is a shared-nothing cluster. Shared-nothing provides high-scalability but requires careful data partitioning. Since OLAP databases are very large and mostly read-only, data partitioning and parallel query processing are effective. However, it is much harder to support OLTP on shared-nothing because of ACID guarantees which require complex concurrency control. For these reasons and because OLTP databases are not so large, shared-disk is the preferred architecture for OLTP.

The second reason that OLTP is not so suitable for cloud is that highly sensitive data get stored at an untrusted host (the provider site). Storing corporate data at an untrusted third-party, even with a carefully negotiated Service Level Agreement (SLA) with a reliable provider, creates resistance from some customers because of security issues. However, this resistance is much reduced for historical data, with anonymized sensitive data.

There is much more variety in cloud data than in scientific data since there are many different kinds of customers (individuals, SME, large corporations, etc.). However, we can identify common features. Cloud data can be very large, unstructured (e.g. text-based) or semi-structured, and typically append-only (with rare updates). And cloud users and application developers may be in high numbers, but not DBMS experts.

## 4 Emerging Solutions

Generic data management solutions (e.g. relational DBMS) that have proved effective in many application domains (e.g. business transactions) are not efficient at dealing with these emerging applications, thereby forcing developers to build ad-hoc solutions that are labor-intensive and cannot scale. In particular, relational DBMSs have been lately criticized for their "one size fits all" approach. Although they have been able to integrate support for all kinds of data (e.g., multimedia objects, XML documents and new functions), this has resulted in a loss of performance and flexibility for applications with specific requirements because they provide both "too much" and "too little". Therefore, it has been argued that more specialized DBMS engines are needed. For instance, column-oriented DBMSs, which store column data together rather than rows in traditional row-oriented relational DBMSs, have been shown to perform more than an order of magnitude better on decision-support workloads. The "one size does not fit all" counter-argument generally applies to cloud data management as well.

Therefore, current data management solutions have traded consistency for scalability, simplicity and flexibility. As alternative to relational DBMS (which use the standard SQL language), these solutions have been recently quoted as Not Only SQL (NOSQL) by the database research community. Many of these solutions have been developed for the cloud or the grid, which both exploit large scale parallelism, typically with shared-nothing clusters.

Figure 3 positions the architectures of emerging solutions along the same three dimensions (distribution, heterogeneity and autonomy). Like cloud computing, grid computing enables access to very large compute and storage resources over the Web. Compared with cloud computing which deals with large-scale parallelism, the grid is characterized with high heterogeneity, large-scale distribution and large-scale parallelism. In addition to grid and cloud, we also position the recent P2P DBMS that target large-scale distribution and data integration systems (like MDBMS) that deal with large-scale distribution, heterogeneity and autonomy.

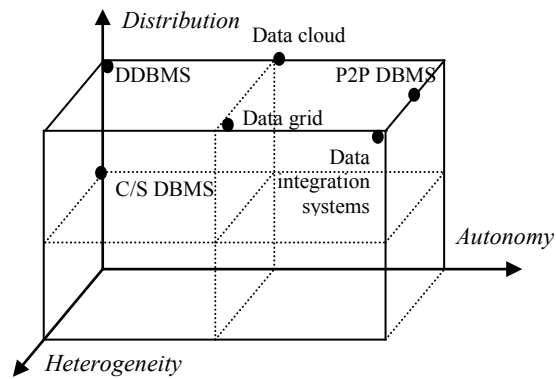

**Figure 3.** Architectures of Emerging Solutions

Distributed data management for cloud applications emphasizes scalability, fault-tolerance and availability, sometimes at the expense of consistency or ease of development. Let us illustrate this approach with two popular solutions: Google Bigtable and Yahoo! PNUTS.

Bigtable is a database storage system for a shared-nothing cluster [4]. It uses a distributed file system (Google File System - GFS) for storing structured data in distributed files, with fault-tolerance and availability. It also uses a form of dynamic data partitioning for scalability. There are also open source implementations of Bigtable, such as Hadoop Hbase, which runs on Hadoop Distributed File System (HDFS). Bigtable supports a simple data model that resembles the relational model, with multi-valued, timestamped attributes. It provides a basic API for defining and manipulating tables, within a programming language such as C++, and various operators to write and update values, and to iterate over subsets of data, produced by a scan operator. There are various ways to restrict the rows, columns and timestamps produced by a scan, as in a relational select operator. However, there is no complex operator such as join or union, which should be programmed using the scan operator. Transactional atomicity is supported for single row updates only. To store a table in GFS, Bigtable uses range partitioning on the row key. Each table is divided into partitions, called tablets, each corresponding to a row range.

PNUTS is a parallel and distributed database system for cloud applications at Yahoo! [5]. It is designed for serving Web applications, which typically do not need complex queries, but require good response time, scalability and high availability and can tolerate relaxed consistency guarantees for replicated data. PNUTS supports the relational data model, with arbitrary structures allowed within attributes of Blob type. Schemas are flexible as new attributes can be added at any time even though the table is being queried or updated, and records need not have values for all attributes. PNUTS provides a simple query language with selection and projection on a single relation. Updates and deletes must specify the primary key. PNUTS provides a replica consistency model that is between strong consistency and eventual consistency, with several API operations with different guarantees. Database tables are horizontally partitioned into tablets, through either range partitioning or hashing, which are distributed across many servers in a cluster (at a site).

To summarize, both Bigtable and PNUTS provide some variations of the relational model, a simple API or language for manipulating data, and relaxed consistency guarantees. They also rely on fragmentation (partitioning) and replication for fault-tolerance. Thus, they capitalize on the well-known principles of distributed data management.

Emerging solutions strive to be more generic than ad-hoc solutions which are labor-intensive and cannot scale. For instance, the SciDB organization (http://www.scidb.org) is building an open source database system for scientific data analytics. SciDB will be certainly effective for similar applications for which the data is well understood (with well-defined data structures). However, to avoid that the "one size fits all" argument applies to SciDB as well, the key question is: *how generic should scientific data management be, without hampering application-specific optimizations?* For instance, to perform scientific data analysis efficiently, scientists typically resort to dedicated indexes, compression techniques and specific algorithms. Thus, generic techniques, inspired from the DB research community should be able to cope with these specific techniques.

Genericity in data management encompasses two dimensions: data model (which provides data structures (captured by the data model) and data processing (inferred by the query language). Relational DBMS have initially provided genericity through the relational data model (that subsumes earlier data models) and a high-level query language (SQL). However, successive object extensions to include new data structures such as lists and arrays and support user-defined functions in a programming language have resulted in a yet generic, but more complex data model and language for the developers. Therefore, emerging solutions tend to rely on a more specific data model (e.g. Bigtable which is some kind of nested relational model) with a simple set of operators easy to use from a programming language. For instance, to address the requirements of social network applications, new solutions rely on a graph data model and graph-based operators. To address the requirements of scientific applications, SciDB supports an array data model, which generalizes the relational model, with array operators. User-defined functions also allow for more specific data processing. MapReduce [6] is a good example of generic parallel data processing framework, on top of a distributed file system (GFS). It supports a simple data model (sets of (key, value) pairs), which allows user-defined functions (map and reduce). Although quite successful among developers, it is relatively low-level and rigid, leading to custom user code that is hard to maintain and reuse. Pig latin [8] is an alternative data management solution that raises the level of abstraction with an algebraic query language. In emerging solutions, it is interesting to witness the development of algebras, with specific operators, to raise the level of abstraction in a way that enables optimization. For instance, in [9], we propose an algebraic approach enables automatic optimization of scientific workflows that manipulate huge amounts of data through specific programs and files.

## 5 Conclusion

To support the requirements of important data-intensive applications (e.g. social networks, web data analytics, scientific applications, etc.), new distributed data management techniques and systems (e.g. MapReduce, Hadoop, SciDB, Peanut, Pig latin, etc.) have emerged and are receiving much attention from the research community. Now, the guiding question for this paper was what is likely to happen in the next decade; or what will be the principles of distributed data management in 2020. I translated this into three questions: (1) What are the fundamental principles behind the emerging solutions? (2) Is there any generic architectural model, to explain those principles? (3) Do we need new foundations to look at data distribution?

To address (1), I illustrated the challenges introduced by new data-intensive applications in the context of scientific applications and cloud computing. The emerging solutions typically provide some variations of the relational model, a simple API or language for manipulating data, and relaxed consistency guarantees. They also rely on fragmentation (partitioning) for large-scale parallelism and replication for fault-tolerance. Thus, they capitalize on the well-known principles of distributed data management.

Wrt (2), I showed that emerging solutions can still be explained along the three main dimensions of distributed data management (distribution, autonomy, heterogeneity), yet pushing the scales of the dimensions high up. However, I raised the question of how generic should distributed data management be, without hampering application-specific optimizations. Emerging NOSQL solutions tend to rely on a specific data model (e.g. Bigtable, MapReduce) with a simple set of operators easy to use from or with a programming language. It is also interesting to witness the development of algebras, with specific operators, to raise the level of abstraction in a way that enables optimization [9]. What is missing to explain the principles of emerging solutions is one or more dimensions on generic/specific data model and data processing.

The hardest question is (3): Do we need new foundations to look at data distribution? It all depends what we mean by "data" and whether we consider the continuum between data, information and knowledge, with increasing pervasiveness of data semantics. During the ICDE 2011 panel [11], one of us (S. Abiteboul) argued that accessing highly distributed, heterogeneous data in personal dataspaces on the web was beyond human expertise, and requires us moving from data to knowledge, with automated reasoning. This is the motivation for the comeback of Datalog as a uniform language to deal with data, metadata, rules, distribution, time, etc. [2].

## Acknowledgements

This paper has benefited from fruitful discussions with many colleagues: Tamer M. T. Özsu, of course, during the writing of [10]; Serge Abiteboul, Bettina Kemme, Ricardo Jiménez-Peris, Beng Chin Ooi in the ICDE 2011 panel on the same topic [11]; the members of the Zenith team at INRIA-LIRMM, in particular, Reza Akbarinia, Florent Masseglia and Esther Pacitti; and the members of the CNPq-INRIA project Datluge, in particular, Marta Mattoso, Eduardo Osgarawa, and Fabio Porto.